\documentclass[aps,prl,twocolumn,longbibliography,nobibnotes,superscriptaddress]{revtex4-2}
\usepackage[english]{babel}
\usepackage{natbib}
\usepackage{ucs}
\usepackage[utf8x]{inputenc}
\usepackage{amsmath}
\usepackage{hyperref}
\usepackage{amsfonts}
\usepackage{amssymb}
\usepackage{graphicx}
\usepackage{bm}
\usepackage{bbm}
\usepackage{braket}

\begin{document}

\title{Tunable Coupling Architecture for Fixed-frequency Transmons}

\author{J.\ Stehlik}
\thanks{These authors contributed equally}
\affiliation{IBM Quantum, IBM T.J.\ Watson Research Center, Yorktown Heights, NY 10598, USA}

\author{D.\ M.\ Zajac}
\thanks{These authors contributed equally}
\affiliation{IBM Quantum, IBM T.J.\ Watson Research Center, Yorktown Heights, NY 10598, USA}

\author{D.\ L.\ Underwood}
\affiliation{IBM Quantum, IBM T.J.\ Watson Research Center, Yorktown Heights, NY 10598, USA}
\author{T.\ Phung}
\affiliation{IBM Quantum, IBM Almaden Research Center, San Jose, CA 95120, USA}
\author{J. Blair}
\affiliation{IBM Quantum, IBM T.J.\ Watson Research Center, Yorktown Heights, NY 10598, USA}
\author{S.\ Carnevale}
\affiliation{IBM Quantum, IBM T.J.\ Watson Research Center, Yorktown Heights, NY 10598, USA}
\author{D.\ Klaus}
\affiliation{IBM Quantum, IBM T.J.\ Watson Research Center, Yorktown Heights, NY 10598, USA}
\author{G.\ A.\ Keefe}
\affiliation{IBM Quantum, IBM T.J.\ Watson Research Center, Yorktown Heights, NY 10598, USA}
\author{A.\ Carniol}
\affiliation{IBM Quantum, IBM T.J.\ Watson Research Center, Yorktown Heights, NY 10598, USA}
\author{M.\ Kumph}
\affiliation{IBM Quantum, IBM T.J.\ Watson Research Center, Yorktown Heights, NY 10598, USA}
\author{Matthias Steffen}
\affiliation{IBM Quantum, IBM T.J.\ Watson Research Center, Yorktown Heights, NY 10598, USA}
\author{O.\ E.\ Dial}
\affiliation{IBM Quantum, IBM T.J.\ Watson Research Center, Yorktown Heights, NY 10598, USA}

\begin{abstract}

Implementation of high-fidelity two-qubit operations is a key ingredient for scalable quantum error correction. In superconducting qubit architectures tunable buses have been explored as a means to higher fidelity gates. However, these buses introduce new pathways for leakage. Here we present a modified tunable bus architecture appropriate for fixed-frequency qubits in which the adiabaticity restrictions on gate speed are reduced. We characterize this coupler on a range of two-qubit devices achieving a maximum gate fidelity of $99.85\%$.  We further show the calibration is stable over one day.

\end{abstract}

\maketitle

Achieving high-fidelity two-qubit (2Q) gates is one of the largest obstacles toward fault-tolerant quantum computation. Many approaches have been developed based on either fixed-frequency or tunable transmons. Tunable transmons naturally allow for fast iSWAP gates between the $\ket{10}$ and $\ket{01}$ states, or controlled-Z (CZ) gates by utilizing the interaction between the $\ket{11}$ and $\ket{20}$ (or $\ket{02}$) states \cite{DiCarlo09,blais04}. However, qubit coherence and stability generally suffer from the presence of flux noise. Alternatively, fixed-frequency qubits offer long coherence times and stability, and a variety of microwave-activated entangling gates can be utilized for two qubit operations \cite{Paraoanu06,Rigetti10,Puri16,Paik16,Chow13}. The drawback to these schemes is that qubit frequencies cannot be tuned away from collisions, and gate times tend to be long \cite{Hertzberg20}. Additionally, regardless of whether using fixed-frequency or tunable transmons, the presence of always-on coupling leads to gate errors caused by spectator qubits \cite{krinner20,mckay19,Cai2021}.

Recently many groups have turned to tunable buses as a way to both achieve faster 2Q gates, and to address issues raised by the presence of always-on coupling \cite{Yan18,Mundada19,Foxen20}. 
Here we explore the tunable bus architecture in a novel regime: with the frequency of the bus below the frequency of the qubits. Here the high ZZ region is near the bus sweet spot, and there is a smooth transition between the on and off locations.  We show that this allows us to achieve high-fidelity gates between fixed frequency transmons.
Our study covers 11 two-qubit devices with varying detunings and bus-qubit coupling strengths. The highest-fidelity gate we achieve is 46~ns long with an error-per-gate (EPG) of $0.0015 \pm 0.0001$, and despite the flux-sensitive nature of the coupler, we find that its optimal calibration is stable for more than one day.

\begin{figure*}
	\centering
	\includegraphics[width=2\columnwidth]{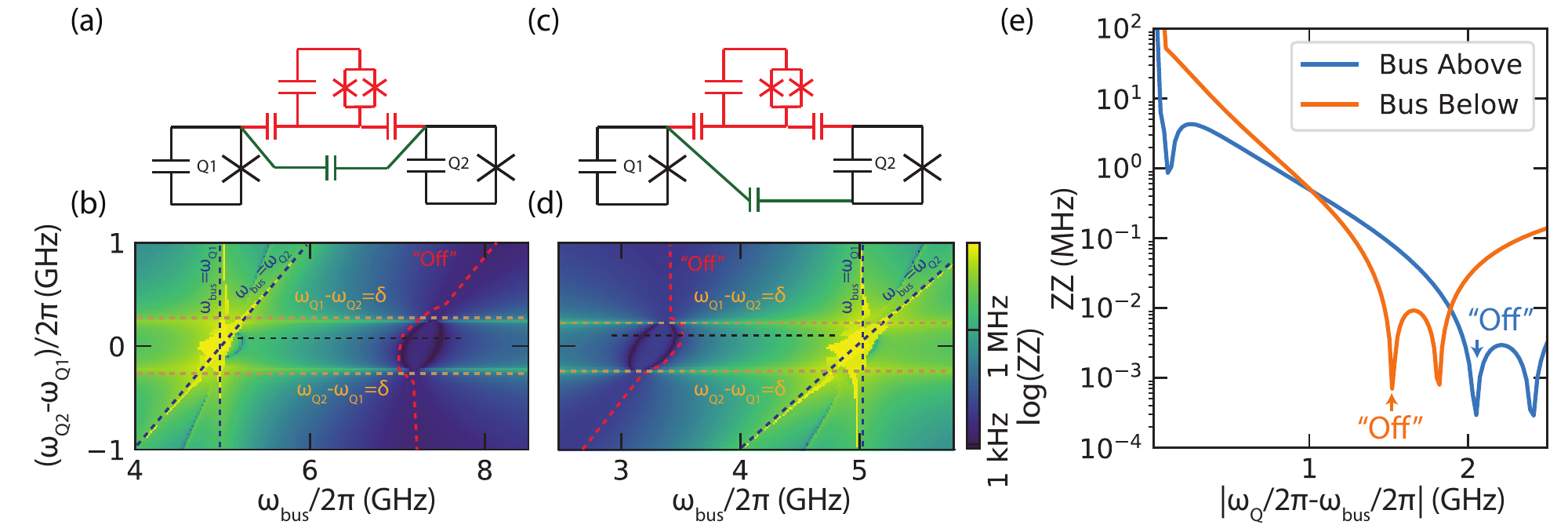}
	\caption{(a) Schematic representation of previously proposed bus above qubits (BAQ) tunable coupler. (b) ZZ as a function of qubit-qubit detuning and bus frequency for BAQ device. (c) Schematic representation of a bus below qubits (BBQ) tunable coupler.  Changing the bypass capacitor to go between opposite islands of the transmons compared to the tunable coupler.  (d) ZZ as a function of qubit-qubit detuning and bus frequency for BBQ device. Due to the change in the bypass capacitor, the zero in ZZ now occurs with the bus detuned below the qubits. 
		(e) ZZ as a function of bus-qubit detuning for both BAQ and BBQ.  For the same magnitude of coupling, BBQ gives larger ``on" coupling and a less complex response curve. }
	\label{fig:intro}
\end{figure*}

In general, achieving a high on-off ratio with a tunable frequency bus is challenging because of the finite tuning range.  For example, to have idle errors below $10^{-4}$ during a 2Q gate, the residual coupling needs to be more than 100 times smaller than the coupling during the on state, which for a simple resonant bus requires increasing the bus detuning 100-fold compared to the device operating point during gates.  To address this shortcoming, F.~Yan~\emph{et al.}\ have proposed a tunable bus scheme \cite{Yan18}, schematically captured in Fig.~\ref{fig:intro}(a).  In this design, a direct capacitor is added in parallel to the tunable bus, and is engineered to cancel the coupling between the two qubits exactly at a particular frequency for the tunable bus.  We can model this system using the following Hamiltonian:
\begin{equation}
H_{\rm tot} = \sum H_{i} + H_{c}
\end{equation}
\begin{equation}
H_{i} = \hbar \omega_{i} a_i^{\dagger} a_i + \frac{\delta_i}{2} a^\dagger_i a^\dagger_i a_i a_i
\end{equation}
where $\omega_i$ is the frequency of the $i$-th transmon with $i \in \{\mathrm{Q1},\mathrm{Q2},\mathrm{bus}\} $, $\delta_i$ is the anharmonicity, and $a^\dagger_i$ ($a_i$) are the raising (lowering) operators.
The coupling Hamiltonian is given by
\begin{equation}
H_{\rm c} = \sum_{i \neq j} g_{i,j} \left( a^\dagger_i + a_i\right) \left( a^\dagger_j + a_j\right). 
\end{equation}

With the bus far-detuned from qubits, we can approximate the effective coupling due to the tunable bus as $J_{\rm bus} \approx \frac{g_{Q1,bus} g_{Q2,bus}}{\Delta}$, where $1 / \Delta = (1 / (\omega_{\rm Q1} - \omega_{\rm bus})+ 1/ (\omega_{\rm Q2} - \omega_{\rm bus}))$.
The direct capacitor between Q1 and Q2 adds a positive coupling term $g_{\rm Q1, Q2}$;  thus, the total effective coupling then becomes $J_{\rm tot} \approx J_{\rm bus} + g_{\rm Q1, Q2}$ \cite{Yan18}.  When $\omega_{\rm bus} > \omega_{\rm Q1,Q2}$, the farther the tunable bus is detuned from the qubits, the closer to zero the (negative) effective $J_{\rm bus}$ is.  Therefore, for any positive small $g_{\rm Q1, Q2}$, it is always possible to find a zero in $J_{\rm tot}$ by adjusting $\omega_{\rm bus}$.  A CZ gate can then be realized by decreasing $\omega_{\rm bus}$ frequency from the ``off" position ($J_{\rm tot} \sim 0$) to an ``on" ($\left|J_{\rm tot}\right| > 0$) position.  Since in this device the tunable bus is detuned above the frequencies of the qubits, for the purposes of the following discussion we will call this the bus above qubits (BAQ) architecture.

We can more quantitatively analyze the system by setting $\delta_{Q1}/2 \pi = \delta_{Q2}/2 \pi = -240$ MHz for the qubits and $\delta_{\rm bus}/ 2 \pi = -140$ MHz for the bus, and using coupling $g_{Q1,bus} / 2 \pi = g_{Q2,bus} / 2 \pi = 110$ MHz and $g_{Q1,Q2}/ 2 \pi = 6$ MHz. 
With these parameters we diagonalize the system and extract ZZ as
\begin{equation}
ZZ = \left( E_{\rm 11;0} - E_{\rm 01;0} - E_{\rm 10;0}  + E_{\rm 00;0} \right) / h.
\end{equation}
Here $E_{\rm nm;l}$ is the energy of the level with $n$ excitations in qubit 1, m excitations in qubit 2, and l excitation in the bus.  We plot ZZ as a function of $\omega_{2}$ and $\omega_{\rm bus}$ in Fig.~\ref{fig:intro}(b) with fixed $\omega_{1} / \left( 2 \pi \right) = 5 $ GHz. 

The plot features four level crossings that result in large ZZ.  We can identify these as the bus being resonant with one of the qubits (either $\omega_{\rm Q1} = \omega_{\rm bus}$ or $\omega_{\rm Q2} = \omega_{\rm bus}$), and with the qubits being detuned by exactly the anharmonicity ($|\omega_{\rm Q1} - \omega_{\rm Q2}|= |\delta|$).  Beyond these regions of high ZZ, the plot shows a zero in ZZ (highlighted using a red dashed line in Fig~\ref{fig:intro}(b) near 7 GHz).  This is surrounded by a broad range of small ZZ, which forms the idle point of the bus.  To form a CZ gate, the bus will be detuned closer to the qubits.

The BAQ design, however, suffers from several drawbacks that make it challenging for adiabatic gates between fixed-frequency transmons, where the detuning between the qubits is not adjustable. Firstly, during gate operation the bus must be detuned below its maximum frequency.  As result, the ``on" state cannot be near the upper sweet spot of the superconducting quantum interference device (SQUID) that forms the tunable bus.  While using an asymmetric SQUID can help, the lower sweet spot is narrower \cite{PhysRevApplied.8.044003}.  This has the effect of increasing sensitivity to noise and also require more precise control of the SQUID critical currents and overall device parameters.

More importantly, if the two qubits are in the straddling regime (i.e., detuned by less than the anharmonicity), there is a significant dip in ZZ that appears with the bus detuned from qubits by $\approx100$ MHz.  We identify this dip with the collision between $E_{\rm 11;0}$ and $E_{\rm 00;2}$ energy levels, that is between the $\ket{2}$ state of the coupler and the $\ket{11}$ state of the two qubits.  This is a two-photon transition that is allowed in the Hamiltonian, and leads directly to a reduction in ZZ compared to what would be seen with a idealized two-level coupler.  Furthermore, due to the two-photon nature of this transition, the associated anti-crossing is in general smaller than the qubit-bus coupling $g_{\rm Q,bus}$.  The presence of this anti-crossing will complicate the dynamics during gate operation, since accessing high-ZZ regions requires passing through the anti-crossing, leading to leakage into the $\ket{2}$ state of bus \cite{Xu2020}.  These dynamics can be avoided by either going to larger detuning $| \omega_{\rm Q1} - \omega_{\rm Q2} | > | \delta_{\rm Q}|$ \cite{Xu2020,Collodo2020} or if the qubits can be dynamically detuned to $\approx 1 \delta$ \cite{Foxen20}.

For these reasons, it is desirable for the off position of the coupler to be such that $\omega_{\rm bus}  < \omega_{\rm qubits}$ -- an alternative bus-below-qubits (BBQ) architecture.   This can be achieved by flipping the sign of any of the $g_{i,j}$.  In our implementation we change the sign of $g_{Q1,Q2}$ by using floating transmons and coupling the islands of the same voltage polarity to the tunable bus, while coupling islands of opposite voltage polarity using the bypass capacitor, as schematically captured in Fig.~\ref{fig:intro}(c). 

If we set $g_{\rm Q1,\rm Q2} / 2 \pi = -6$~MHz and keep all other parameters the same as our BAQ simulation, the resulting plot of ZZ as a function of detuning and bus frequency is depicted in  Fig.~\ref{fig:intro}(d).  We again have a region of low ZZ, which in this case occurs with the bus near $3$ GHz. The decrease associated with the $\ket{2}$ state of the bus is still present, but in this case does not interfere with turning on the gate because it is above the lower qubit frequency, and above the intended operating point of the bus.

To more quantitatively compare BAQ and BBQ devices, we fix the detuning at 90 MHz and plot ZZ as a function of the detuning between the bus and the qubit closer in frequency to the bus.  This is plotted in Fig.~\ref{fig:intro}(e).  We see that for the same absolute magnitude of coupling and bus detuning, we can achieve more ZZ contrast in the BBQ architecture.  In BBQ we can also turn on gates ZZ exceeding $10$ MHz without going through any avoided crossings (for example, 11;0 and 00;2 -- which manifests as a dip in ZZ for the BAQ device).  As a result, adiabaticity constraints are loosened in the BBQ architecture; this allows for adiabatic CZ gates, which are faster and suffer from less leakage. This makes BBQ architecture advantageous for gates between fixed-frequency transmons.

\begin{figure}
	\centering
	\includegraphics[width=1\columnwidth]{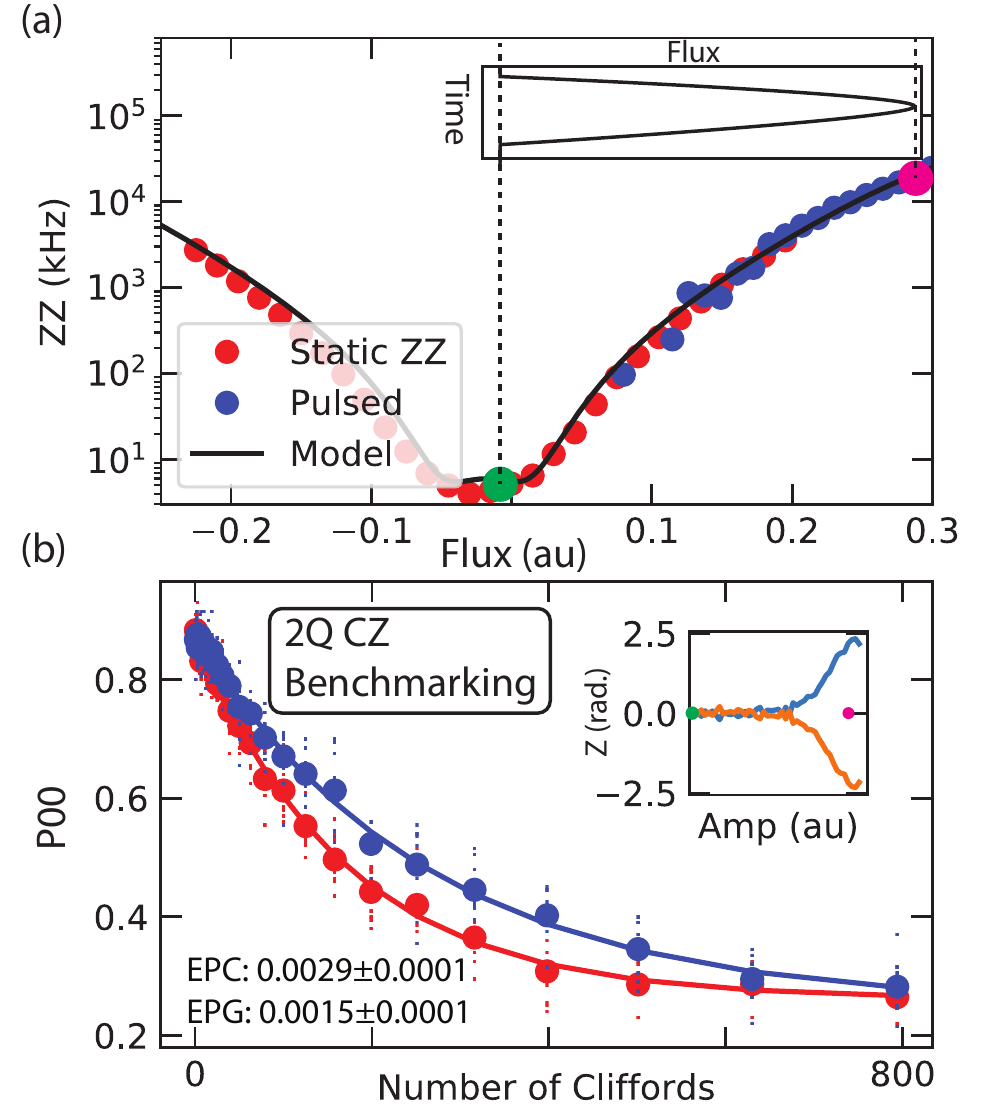}
	\caption{(a) ZZ as a function of applied bias measured on qubit pair 1.  To create a two-qubit gate, we apply static bias to minimize ZZ and then apply pulse to a region of high ZZ. Superimposed with the time axis vertically is a schematic rendition of the applied flux pulse.  (b) Two-qubit randomized benchmarking: probability of measuring qubits in the $\ket{00}$ state as a function of number of Cliffords.  We measure an EPG of 0.0015.  Inset shows Z rotation as a function of pulse amplitude for two states of the control qubit.}
	\label{fig:gate_basics}
\end{figure}

\begin{figure}
	\centering
	\includegraphics[width=1\columnwidth]{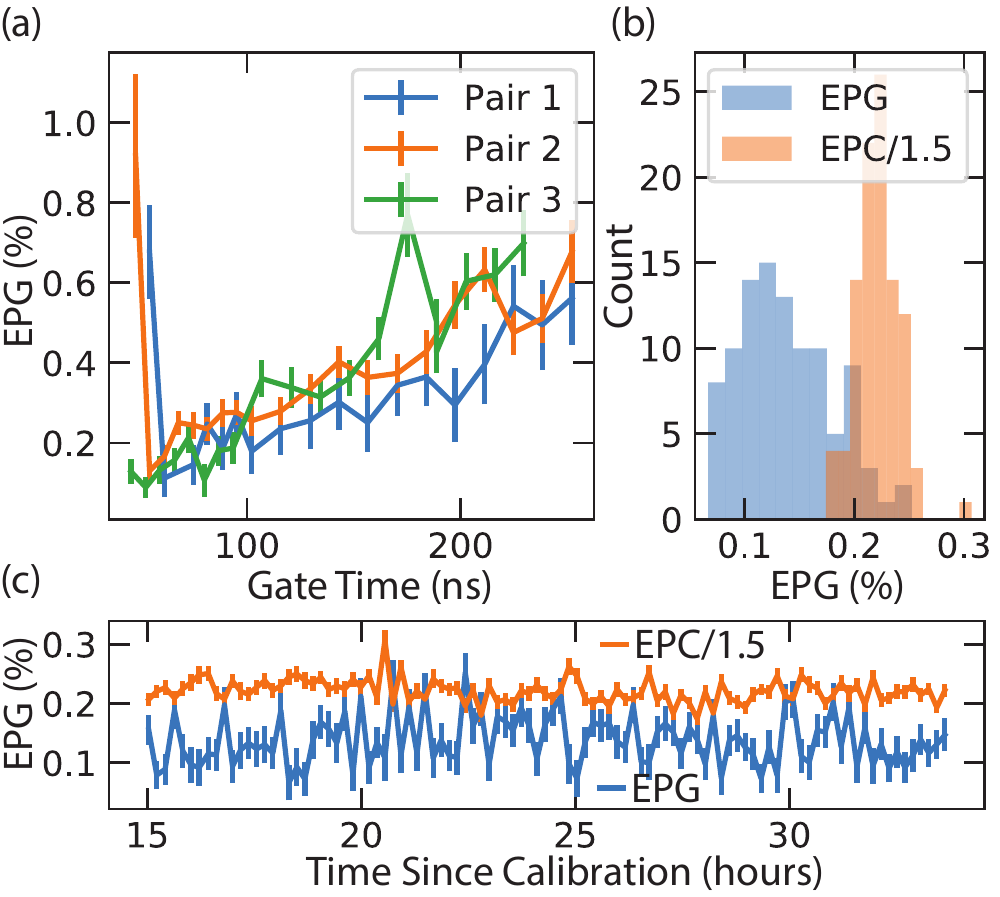}
	\caption{(a) EPG by interleaved benchmarking as a function of gate length for three 2Q devices. (b)  Histogram of EPG and EPC/1.5 measured over a period of 20 hours. (c) EPG vs.\ time since last calibration on the gate from (b), demonstrating good stability for at least 24 hours, despite the use of flux-tunable elements. }
	\label{fig:best_gate}
\end{figure}

To study and validate the BBQ architecture experimentally, we fabricated several 2Q devices with varying coupling parameters.  To enable high-fidelity gate operations, we use an asymmetric SQUID to deliberately limit the tunability of the coupler qubit to approximately the range needed to both achieve the minimum ZZ and reach $\sim 50$~MHz ZZ.  In Fig.~\ref{fig:gate_basics}(a) we plot measured ZZ as a function of the bias applied to the SQUID loop for a typical device with qubit-qubit detuning of 351~MHz.  In regions of high ZZ, where it is impossible to calibrate single-qubit gates, we instead rely on a pulsed measurement.  We fit the observed ZZ vs.\ flux curve and obtain approximately $g_{Q1,bus} / 2 \pi \approx 130$~MHz, $g_{Q2,bus} / 2 \pi \approx 120$~MHz, and $g_{Q1,Q2} / 2 \pi \approx -4$~MHz, which is approximately 10\% larger than our designed values.

To operate the device, we bias to a point of low ZZ.  For the parameters chosen we achieve $\text{ZZ} < 10$~kHz with the coupler in its off state.  To implement a two~qubit gate, we can now apply a pulse to adiabatically shift the device into a configuration with high ZZ, accumulate a phase difference, and then pulse back. This is schematically captured in the inset of Fig.~\ref{fig:gate_basics}(a).

Pulse shaping is required to maintain adiabaticity. To derive our final flux pulse shape, we use a technique similar to one described in Ref.~\cite{Martinis2014}; see \cite{SOM}.  We calibrate our pulses by putting one qubit (Q1) on the equator, and preparing the other qubit (Q2) in either the $\ket{1}$ or the $\ket{0}$ state.  We then apply pulses of varying amplitude.  Due to the ZZ interaction, depending on the state of Q2, Q1 will accumulate a different phase. This is shown in the inset of Fig.~\ref{fig:gate_basics}(b), where we plot the Z rotation for the $\ket{1}$ and the $\ket{0}$ state of Q2.  We choose an amplitude that results in a difference of $\pi$ between the $\ket{0}$ and $\ket{1}$ rotations, and apply a software Z-rotation \cite{McKay2017} to both qubits to achieve the following gate unitary:
\begin{equation}
U_{\rm CZ} =
\begin{pmatrix}
1 & 0 & 0 & 0 \\
0 & 1 & 0 & 0 \\
0 & 0 & 1 & 0 \\
0 & 0 & 0 & -1. 
\end{pmatrix}
\end{equation}

We perform interleaved randomized benchmarking on this gate and obtain an EPG of $0.0015 \pm 0.0001$ \cite{Corcoles2013}.  Crucially, we can achieve this error without sacrificing single-qubit gate fidelity, and as a result the non-interleaved benchmarking decay is visible to 800 Cliffords.  By dividing the observed error-per-Clifford (EPC) by the average number of two-qubit gates in our Clifford decomposition \cite{SOM}, we find the EPG is bounded from above by $0.0019 \pm 0.0001$, which is the gate error we would deduce, assuming our single-qubit gates are perfect.

To ensure reproducibility of these results, we study this gate scheme across different devices.  In Fig.~\ref{fig:best_gate}(a) we plot gate fidelity as extracted from interleaved randomized benchmarking for three different 2Q pairs, as a function of gate length.  All three pairs exhibit similar behavior.  With longer gate times we see an increase in error due to qubit decoherence.  At very short time scales (near 50~ns) we see an increase in error, which we attribute to the evolution during the gate no longer being adiabatic.

One crucial aspect in any gate scheme is stability.  This is especially poignant in a flux-based architecture, with 1/f flux noise ultimately limiting the repeatability \cite{PhysRevA.76.042319}.  In our BBQ architecture, however, the gate ``on" position can be near the upper sweet spot of the SQUID tuning range.  We can use this to realize excellent stability in time.  To demonstrate this, we repeatedly measure gate fidelity over a 19-hour period without recalibration.  In Fig.~\ref{fig:best_gate}(b) we plot the histogram of the fidelities extracted by interleaved randomized benchmarking, as well as the upper bound on error (established by dividing the EPC by the number of two-qubit gates in our decomposition).  Both distributions show excursions in fidelity that are, at worst, 60\% worse.  In Fig.~\ref{fig:best_gate}(c) we plot the same data as a function of time since calibration. Over the entire measurement period, we observe no significant deviations from the quoted fidelity; furthermore, we do not observe any clear drift towards degraded performance.

We also extend the study to cover different detunings and coupling strengths.  A summary of the 11 devices characterized is found in Table~\ref{tab:gates}.  In general, higher coupling strength results in faster and higher-fidelity gates.  While only 1 pair with a design of $g_{Q,bus}/ 2\pi= 80$~MHz achieved error less than 0.002, such error was achieved on three of the five pairs with a design of $g_{Q,bus} / 2 \pi = 110$~MHz.

\begin{table*}
	
	\caption{Summary of errors and coherences over eleven qubit pairs studied. }
	\label{tab:gates}
	\begin{tabular}{|c|c|c|c|c|c|c| c|}

		\hline
		\multicolumn{1}{|p{1.5cm}|}{\centering Qubit \\ Pair}
		& \multicolumn{1}{|p{1.5cm}|}{\centering Coupling \\ (MHz)}
		& \multicolumn{1}{|p{1.5cm}|}{\centering Detuning \\ (MHz)}
		& \multicolumn{1}{|p{2cm}|}{\centering Average $T_1$ \\ ($\mu$s)}
		& \multicolumn{1}{|p{2cm}|}{\centering Average $T_2$ \\ ($\mu$s)}
		& \multicolumn{1}{|p{1.75cm}|}{\centering Gate Time \\ (ns)}
		& \multicolumn{1}{|p{1.75cm}|}{\centering Error Per \\ Gate}
		& \multicolumn{1}{|p{1.75cm}|}{\centering Error Per \\ Clifford} \\
		\hline
		1 & 110 & 351 &  76 & 105 & 46 & $0.0015\pm0.0001$ & $0.0029\pm0.0001$ \\
		2 & 110 & 398 &  82 & 151 & 70 & $0.0019\pm0.0002$ & $0.0039\pm0.0003$ \\

		3 & 110 & 324 &  68 & 134 & 54 & $0.0018\pm0.0001$ & $0.0037\pm0.0002$ \\
		
		4 & 110 & 503 &  77 & 104 & 130 & $0.0056\pm0.0003$ & $0.0094\pm0.0004$ \\

		5 & 110 & 562 &  86 & 113 & 81 & $0.0025\pm0.0002$ & $0.0048\pm0.0002$ \\

		6 & 80 & 136 &  100 & 124 & 63 & $0.0020\pm0.0004$ & $0.0044\pm0.0002$ \\
		7 & 80 & 145 &  70 & 77 & 86 & $0.0035\pm0.0004$ & $0.0055\pm0.0003$ \\
		8 & 80 & 9 &  80 & 107 & 136 & $0.006\pm0.002$ & $0.011\pm0.001$ \\
		9 & 80 & 166 &  46 & 106 & 113 & $0.0046\pm0.0007$ & $0.0088\pm0.0004$ \\
		10 & 80 & 89 &  63 & 80 & 56 & $0.005\pm0.002$ & $0.007\pm0.001$ \\
		11 & 80 & 160 &  24 & 35 & 130 & $0.0049\pm0.0006$ & $0.0085\pm0.0003$ \\
		
		\hline
	\end{tabular}
\end{table*}

One noteworthy entry in Table~\ref{tab:gates} is pair 8.  This pair had a detuning of just 9 MHz. Such low detuning poses a significant problem for our chosen adiabatic CZ gate.  
This is due to the fact that moving the bus towards the qubits generates an effective exchange interaction between the qubits.  The turn on of this exchange interaction needs to be done adiabatically with respect to the detuning between the qubits, otherwise a partial swap will be generated between the qubits.  Such swaps have been used to generate arbitrary iSWAP and CZ angles in architectures with tunable qubits \cite{Foxen20}.

For qubit pair 8, we cancel the swap operation without detuning the qubits by employing a technique similar to sudden net zero scheme in Ref.~\cite{Negirneac2020}.  We use the fact that a single pulse can be thought of as a beam splitter -- splitting between the swapped and not-swapped possibilities.  A second pulse then forms an interferometer and will either constructively or destructively interfere, depending on the path length difference.  In our case we adjust the path length by varying the timing between the pulses.   We thus split our gate pulse into two.  We use the amplitude of the pulses to adjust the CZ angle and we adjust the wait time between the pulses to cancel out the iSWAP angle.

\begin{figure}
	\centering
	\includegraphics[width=1\columnwidth]{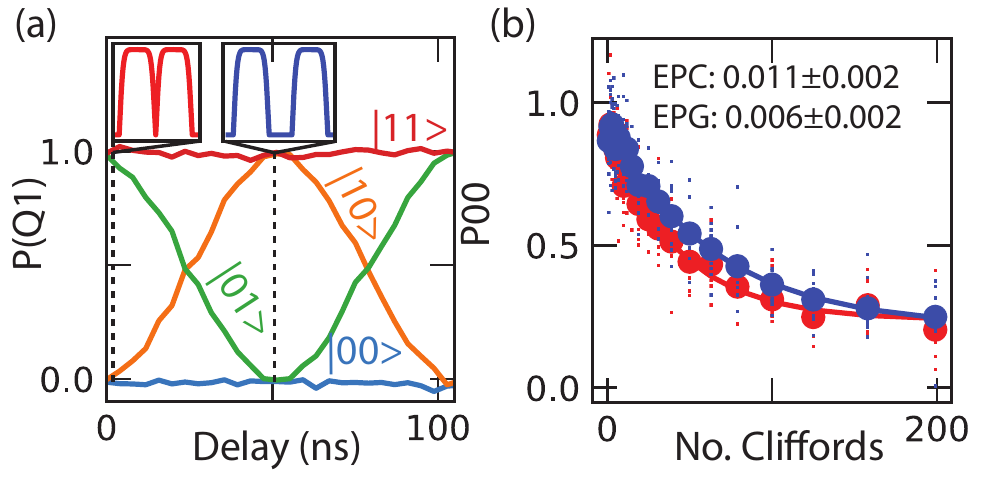}
	\caption{(a) Probability of measuring the first qubit in the excited state as function of the wait-time between pulses for input states of $\ket{00}$, $\ket{01}$, $\ket{10}$, and $\ket{11}$.  Depending on the delay, a iSWAP or a CZ gate can be generated. (b) Probability of finding the qubits in a $\ket{00}$ state as a function of number of applied Cliffords.  We extract an EPC of 1$\%$ for our highest fidelity double pulsed gate.}
	\label{fig:double_pulse}
\end{figure}

We explore the swap mechanism in Fig.~\ref{fig:double_pulse}(a) where we plot the probability of finding Q1 in the excited state as a function of the wait time for the 4 basis states $\ket{00}$, $\ket{10}$, $\ket{01}$, and $\ket{11}$.  While with no delay the pulse sequence results in an almost perfect swap, at 50 ns the swap is canceled.  
In Fig.~\ref{fig:double_pulse}(b) we plot the results of interleaved randomized benchmarking. Our double pulsed gate of 110 ns achieves an EPG of $0.006\pm0.002$.  Due a high readout error on this pair we implement basic readout correction described in Ref.~\cite{watson2018programmable} for the data presented in Fig.~\ref{fig:double_pulse}(b).

In conclusion, we developed a new type of a tunable coupler that is suitable for operations between fixed-frequency transmons.  Using this BBQ architecture we demonstrated gate fidelities of 99.85\% on isolated two-qubit devices.  We obtained similar results on additional devices, showing the repeatability of this scheme.  BBQ also allows us to operate with the coupler near its flux sweet spot, resulting in enhanced stability.

Acknowledgments: We thank D.\ McKay for helpful discussions, P.\ Gumann and J.\ Rozen for  cryostat assistance, and A.\ C\'{o}rcoles, M.\ Takita, A.\ Finck, and A.\ Kandala for technical assistance during the quarantine.

\bibliography{references}

\widetext
\pagebreak

\begin{center}
	\textbf{\large Supplement to Tunable Coupling Architecture for Fixed Frequency Transmons}
\end{center}
\setcounter{equation}{0}
\setcounter{figure}{0}
\setcounter{table}{0}
\setcounter{page}{1}
\makeatletter
\renewcommand{\theequation}{S\arabic{equation}}
\renewcommand{\thefigure}{S\arabic{figure}}
\renewcommand{\bibnumfmt}[1]{[S#1]}

\section{Pulse Shaping}

When the two qubits are detuned far from any frequency collision one significant source of error is leakage to the bus. We thus want the pulse to be as adiabatic with respect the anti-crossing between the bus and the lower qubit as possible.  We can derive a simple pulse shape that accomplishes this goal by considering the 1-photon subspace in isolation.  Here the photon either lives in the qubit or in the bus and the dynamics are governed by the following Hamiltonian:

\begin{equation}
H =
\begin{pmatrix}
\Delta_{bus} & g_{\rm Q1,\rm bus}  \\
g_{\rm Q1,\rm bus}  & 0 
\end{pmatrix}
\end{equation}

Here $\Delta_{bus} = \omega_{Q1} - \omega_{bus}$.  Using this Hamiltonian we can define quantization angle as:
\begin{equation}
\theta = \tan^{-1} \frac{\Delta_{\rm bus}}{2 g_{\rm Q1, \rm bus}}.
\end{equation}
We can think of this angle as defining the precession axis on a Bloch sphere of a fictitious qubit spanned by the two states.  The state vector of the two-level system will precess around this quantization axis.  Thus for an adaiabatic pulse it is desirable to change the quantization angle as slowly as possible.  For a pulse of fixed length this means $d \theta / dt$ should be constant.    This is impractical for a pulse that starts and ends at the same $\Delta_{bus}$ as that would produce a discontinuity in $d \theta / dt$.  Instead we can create a pulse where $d \theta / dt = c \tan^{-1} \left(a t \right)$, where $a$ parametrizes how fast the pulse turns around $c$ is chosen such that pulse spans the correct range of $\theta$.  Integrating $d \theta / dt$ and solving for the integration constant and $c$, we can derive a pulse shape that will start and end at $\Delta_{\rm bus} = \Delta_1$ at $t = -l$ and $t= l$ respectively, while reaching a maximum of $\Delta_{\rm bus} = \Delta_2$ at $t = 0$ as:

\begin{equation}
\Delta_{\rm bus}(t) = 
2 g \tan \left(\frac{2 \left(a \tan
	^{-1}\left(\frac{\Delta _1}{2 g}\right)-a \tan
	^{-1}\left(\frac{\Delta _2}{2 g}\right)\right)
	\left(t \tan ^{-1}(a t)-\frac{\log \left(a^2
		t^2+1\right)}{2 a}\right)}{2 a l \tan ^{-1}(a
	l)-\log \left(a^2 l^2+1\right)}+\tan
^{-1}\left(\frac{\Delta _2}{2
	g}\right)\right)+\Delta _1.
\end{equation}
In practice we have obtained better results by using larger than expected $g$ as a parameter for our pulse.  This is likely due to the fact that the above derivation is only an approximation and ignores other levels.

\section{Clifford Decomposition}
\label{sec:benchmarking}

To implement randomized benchmarking we use the following set of single qubit gates: $X_{\pi}$, $X_{\pi /2}$, $Z_{\pm \pi}$, $Z_{\pm \pi/2}$ and the $CZ$ two qubit gate.  $Z$ gates are implemented using a phase shift on the following gates, following Ref.~\cite{McKay2017}.  

Due to incomplete optimization devices with 80 MHz coupling used on average 1.87 CZ gates per Clifford and 3.17 non-Z single qubit gates.  For devices with 110 MHz coupling we further optimized the Clifford decomposition and used on average 1.5 CZ gates per Clifford and 3.26 non-Z single qubit gates.

\end{document}